\documentclass[prx,superscriptaddres, twocolumn]{revtex4-2}

\usepackage{amsmath, amsfonts,amssymb,amsthm, graphicx, tikz, booktabs, bm}
\usepackage{graphicx}% Include figure files
\usepackage{dcolumn}% Align table columns on decimal point
\usepackage{bm}% bold math       % images
\usepackage{float} 
\usepackage[colorlinks=true,
            linkcolor=blue,
            citecolor=blue,
            urlcolor=blue]{hyperref}
\DeclareGraphicsExtensions{.pdf,.png,.jpg}

\begin{document}

\title{Thermodynamic connectivity reveals functional specialization and multiplex organization of extrasynaptic signaling}
\author{Giridhar Sunil}
% \email[Corresponding author: ]{giridhar.sunil@concordia.ca}
\affiliation{Department of Electrical and Computer Engineering, Concordia University, Montreal, QC, H3G 1M8}
\affiliation{Centre de Recherche de l'Institut Universitaire de G\'eriatrie de Montr\'eal (CRIUGM), Montr\'eal, QC, Canada}
\author{Habib Benali}
% \thanks{These authors equally supervised this work.}
% \email[Corresponding author: ]{habib.benali@concordia.ca}
% \altaffiliation{Corresponding author: elkaioum.moutuou@concordia.ca; These authors equally supervised this work.}
\affiliation{Department of Electrical and Computer Engineering, Concordia University, Montreal, QC, H3G 1M8}
\affiliation{Centre de Recherche de l'Institut Universitaire de G\'eriatrie de Montr\'eal (CRIUGM), Montr\'eal, QC, Canada}
% \affiliation{Inserm U1146, Paris, France}
\author{Elka\"ioum M.~Moutuou}
\email[Corresponding author: ]{elkaioum.moutuou@concordia.ca}
\affiliation{Department of Electrical and Computer Engineering, Concordia University, Montreal, QC, H3G 1M8}
\affiliation{Centre de Recherche de l'Institut Universitaire de G\'eriatrie de Montr\'eal (CRIUGM), Montr\'eal, QC, Canada}

\date{\today}

\begin{abstract}
Neural communication operates on both fast synaptic transmission and slower, diffusive
extrasynaptic signaling, yet how these two modes jointly organize brain function remains
unclear. Here, using the complete synaptic and neuropeptidergic connectomes of
\emph{Caenorhabditis elegans}, we develop a unified multiplex framework linking anatomical
wiring to functional communication. We infer structure-derived functional connectivity
from the synaptic connectome using equilibrium principles from statistical physics,
yielding a probabilistic map of information flow across all synaptic pathways, and compare
this functional layer directly with the extrasynaptic connectome. This reveals a principled
functional specialization across four communication regimes: (i) a topology-dependent
layer that reinforces and stabilizes synaptic motor circuits, (ii) a topology-resilient
modulatory layer supporting global regulation and behavioral state control, (iii) a purely
extrasynaptic network sustaining survival and homeostasis, and (iv) a purely synaptic
regime mediating rapid, low-latency sensorimotor processing. Together, these findings
reveal that synaptic and extrasynaptic signaling form complementary architectures
optimized for speed, modulation, robustness, and survival, and provide a general strategy
for integrating structural and modulatory connectomes to understand how distinct
communication modes cooperate to sustain coherent brain function.

\end{abstract}

\maketitle
%\tableofcontents

\medskip 
\textbf{Significance.} Brains communicate through fast synaptic wiring and slower, diffuse extrasynaptic
signaling, but how these modes jointly organize function has remained unclear. By
integrating the synaptic and neuropeptidergic connectomes of \emph{C.~elegans} within a
unified functional framework, we show that extrasynaptic signaling is not merely an
extension of synaptic wiring but forms structured, functionally specialized communication
regimes. Specifically, we identify four complementary regimes: a topology-dependent layer
that reinforces synaptic motor circuits, a topology-resilient layer supporting global
modulation and behavioral state control, a purely extrasynaptic network sustaining
survival and homeostasis, and a purely synaptic regime mediating rapid sensorimotor
responses. These results reveal a principled functional organization of neural
communication and provide a general strategy for integrating structural and modulatory
connectomes --- one that can be extended to other organisms as multimodal connectivity
data become available.

\begin{figure*}[htbp]
  \centering
  \includegraphics[width=0.8\textwidth]{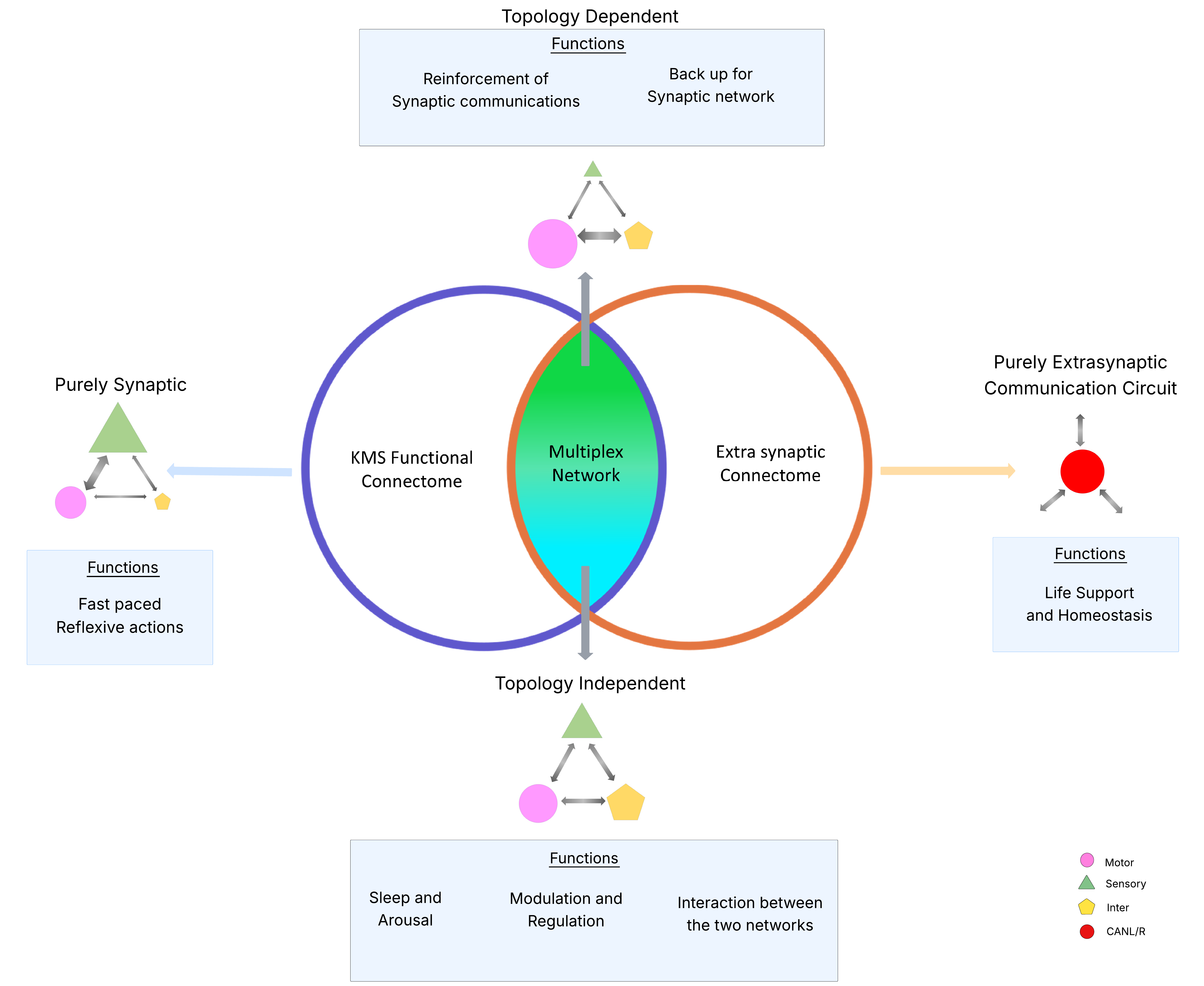}
  \caption{\textbf{Functional specialization across multiplex communication regimes}.
  The multiplex network partitions into distinct communication regimes with complementary
  functional roles. The topology-dependent regime reinforces synaptic circuits,
  particularly in motor control. The topology-resilient regime supports global modulation
  and behavioral state regulation. The purely extrasynaptic regime sustains survival and
  homeostatic functions. In contrast, rapid sensorimotor processing is mediated primarily
  by purely synaptic pathways. Together, these regimes define a principled functional
  organization across signaling modes specialized for speed, modulation, robustness, and
  survival.}
  \label{fig:full_image}
\end{figure*}

\section{\label{sec:intro}Introduction}
Neural systems communicate through multiple signaling modalities operating across distinct
spatial and temporal scales. Fast synaptic transmission enables rapid, point-to-point
communication essential for sensation and motor control, whereas slower extrasynaptic
signaling, mediated by neuropeptides and volume transmission, modulates neural activity
over broader regions and longer timescales~\cite{white1986structure,pereda2014,
ripoll2023neuropeptidergic,watteyne2024neuropeptide,flavell2013serotonin,antal2025molecular,
gilson2019network}. Both modes are essential for
behavior, yet how they jointly organize information flow and functional specialization
remains poorly understood~\cite{bentley2016multilayer, ripoll2023neuropeptidergic, randi2023celegans}.

Although synaptic connectomes have been mapped in increasing resolution, anatomical wiring
alone does not fully determine neural function~\cite{varshney2011structural,cook2019connectome,winding2023insect,Veraszto2024,Scheffer2020}. Neurons can influence distant
targets without direct synaptic contact, and neuromodulatory signals can reshape circuit
dynamics, behavioral states, and physiological regulation~\cite{bentley2016multilayer, Douglas2004}. Extrasynaptic
signaling, in particular, has been implicated in sleep, arousal, stress responses,
feeding, learning, and homeostasis~\cite{ripoll2023neuropeptidergic,flavell2013serotonin, watteyne2024neuropeptide,Barrios2012,atkinson2021ascaris}. These observations reveal that extrasynaptic
communication extends beyond reinforcing synaptic circuits, but whether extrasynaptic
pathways depend on, complement, or operate independently of synaptic wiring remains
unresolved~\cite{bentley2016multilayer}.
% (see Fig.~\ref{fig:multiplex}).

The nematode \emph{Caenorhabditis elegans} provides a unique opportunity to address this
question. Its nervous system comprises a compact and nearly complete synaptic connectome,
alongside a recently mapped neuropeptidergic connectome capturing extrasynaptic signaling
pathways~\cite{cook2019connectome,ripoll2023neuropeptidergic}. This dual connectivity framework enables a systematic investigation
of how distinct communication modes interact within a single brain. However, integrating
these layers requires a functional reference frame that goes beyond structural connectivity
and quantifies how information is likely to propagate through synaptic
networks~\cite{Flavell2022}.

Here we infer structure-derived functional connectivity from the synaptic connectome using
equilibrium principles from statistical physics. Specifically, we apply the
Kubo-Martin-Schwinger (KMS) formalism to model neural interactions as a thermodynamic
system at equilibrium~\cite{moutuou2025brain,moutuou2025kubo}, assigning probabilistic
weights to all synaptic pathways. Embedding this functional synaptic layer within a
multiplex network alongside the extrasynaptic connectome~\cite{ripoll2023neuropeptidergic}
enables a direct comparison between wired and diffusive communication modes.

This approach reveals a principled functional specialization across neural communication
pathways. Extrasynaptic signaling partitions into three regimes: a topology-dependent
layer reinforcing and stabilizing synaptic motor circuits, a topology-resilient modulatory
layer supporting global regulation and behavioral state control, and a purely extrasynaptic
network sustaining survival and homeostasis. In contrast, rapid sensorimotor functions are
mediated primarily by a distinct, purely synaptic communication regime. These four regimes (Fig.~\ref{fig:full_image})
define complementary signaling architectures specialized for speed, modulation, robustness,
and survival, providing a unified view of how distinct communication modes cooperate to
sustain brain function.

\section{\label{sec:Results} Results}
\begin{figure}[!h]
  \centering
  \includegraphics[width=.45\textwidth]{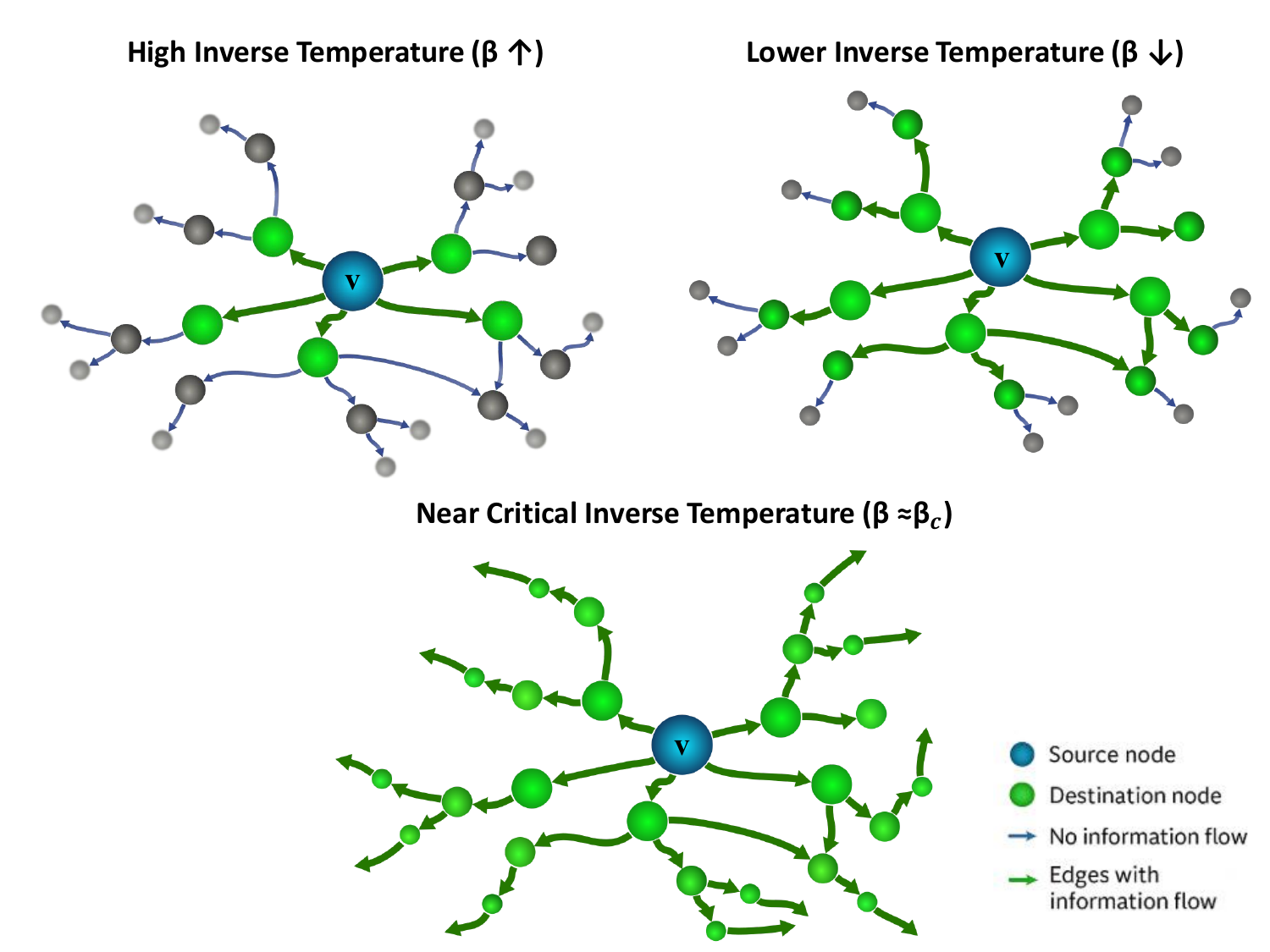}
  \caption{\textbf{Scale-dependent propagation of neural emittance}. KMS-derived emittance profiles illustrate how information flow from a source neuron reorganizes across inverse temperature regimes ($\beta$). At high inverse temperature ($\beta \uparrow$), communication is localized and dominated by direct synaptic connections. As $\beta$ decreases, multi-step pathways increasingly contribute, producing distributed connectivity patterns. Near the critical regime ($\beta \approx \beta_c$), information flow becomes maximally dispersed across the network. Node size reflects distance from the source, and edge coloring distinguishes active from inactive communication pathways. This scale dependence underlies the construction of the structure-informed functional connectome.}
  \label{fig:kms_up}
\end{figure}

% \subsection{Thermal Equilibrium in Directed Multigraphs}
\subsection{Neural Emittance from KMS states}

To compare synaptic and extrasynaptic communication, we first establish a functional
reference frame that captures how information propagates through synaptic wiring beyond
direct anatomical connections. We derive a structure-informed functional connectome (SIFC)
from the \emph{C.~elegans} synaptic connectome using the Kubo-Martin-Schwinger (KMS)
equilibrium framework from algebraic quantum mechanics~\cite{haag1967quantum,
moutuou2025kubo, moutuou2025brain}. The central insight is that a connectome encodes not
only which neurons are directly wired, but also the full space of multi-step pathways
through which signals can propagate. The KMS formalism treats this space as a
thermodynamic system at equilibrium, assigning exponentially decaying weights to paths of
increasing length and thereby converting the static wiring diagram into a probabilistic
map of information flow~\cite{moutuou2025brain,moutuou2025kubo}. This approach captures
the cumulative influence of recurrent loops and indirect routes that are invisible to
pairwise structural measures.

The inverse temperature $\beta$ acts as a propagation scale parameter: at high $\beta$,
communication is dominated by short, direct synaptic paths, whereas as $\beta$ decreases
toward a critical value $\beta_c$, longer and weaker multi-step pathways contribute
increasingly, producing more distributed patterns of functional connectivity
(Fig.~\ref{fig:kms_up}). In all analyses we use $\beta = 5.103$, selected as the value
that maximizes the entropy of the emittance distribution across weighting schemes,
corresponding to the most informationally diverse yet convergent regime. We refer to the Supplementary Information (SI) for more details.

We model the connectome as a directed graph comprising 302 neurons and 16,857 directed
synapses obtained by combining chemical synapses and gap-junction networks~\cite{
varshney2011structural,cook2019connectome,moutuou2025brain}. The KMS construction assigns
to each neuron $v$ an \emph{emittance} $\mathbf{x}^{v|\beta}$: a distribution over all
target neurons $u$ quantifying the probability that a signal originating from $v$ reaches
$u$ under equilibrium propagation. Specifically,
\begin{equation}
\mathbf{x}^{v|\beta}_{u} = \frac{1}{Z^{\beta}_{v}}
\left( \sum_{\mu: v \to u} e^{-\beta|\mu|}\right),
\label{eq:emittance}
\end{equation}
where the sum is over all directed walks $\mu$ of any length from $v$ to $u$, $|\mu|$ is
the number of synapses traversed, and $Z^{\beta}_{v}$ is a normalization factor
representing the total communication capacity of neuron $v$ at inverse temperature $\beta$
(see SI). Because the graph contains cycles and self-loops, equation~\eqref{eq:emittance}
is an infinite series that converges only for $\beta > \beta_c$, where $\beta_c$ is
determined by the spectral radius of the adjacency matrix; in this regime it admits a
compact resolvent form (see SI).

At a fixed $\beta$, the emittance of each neuron is a vector over all targets. Assembling
these column vectors into a matrix --- so that entry $(u,v)$ gives the probability of
information transfer from $v$ to $u$ --- defines the SIFC. This matrix encodes functional
connectivity across all neuron pairs, integrating contributions from both direct synaptic
connections and higher-order network structure.

\begin{figure}[!h]
  \centering
  \includegraphics[width=\linewidth]{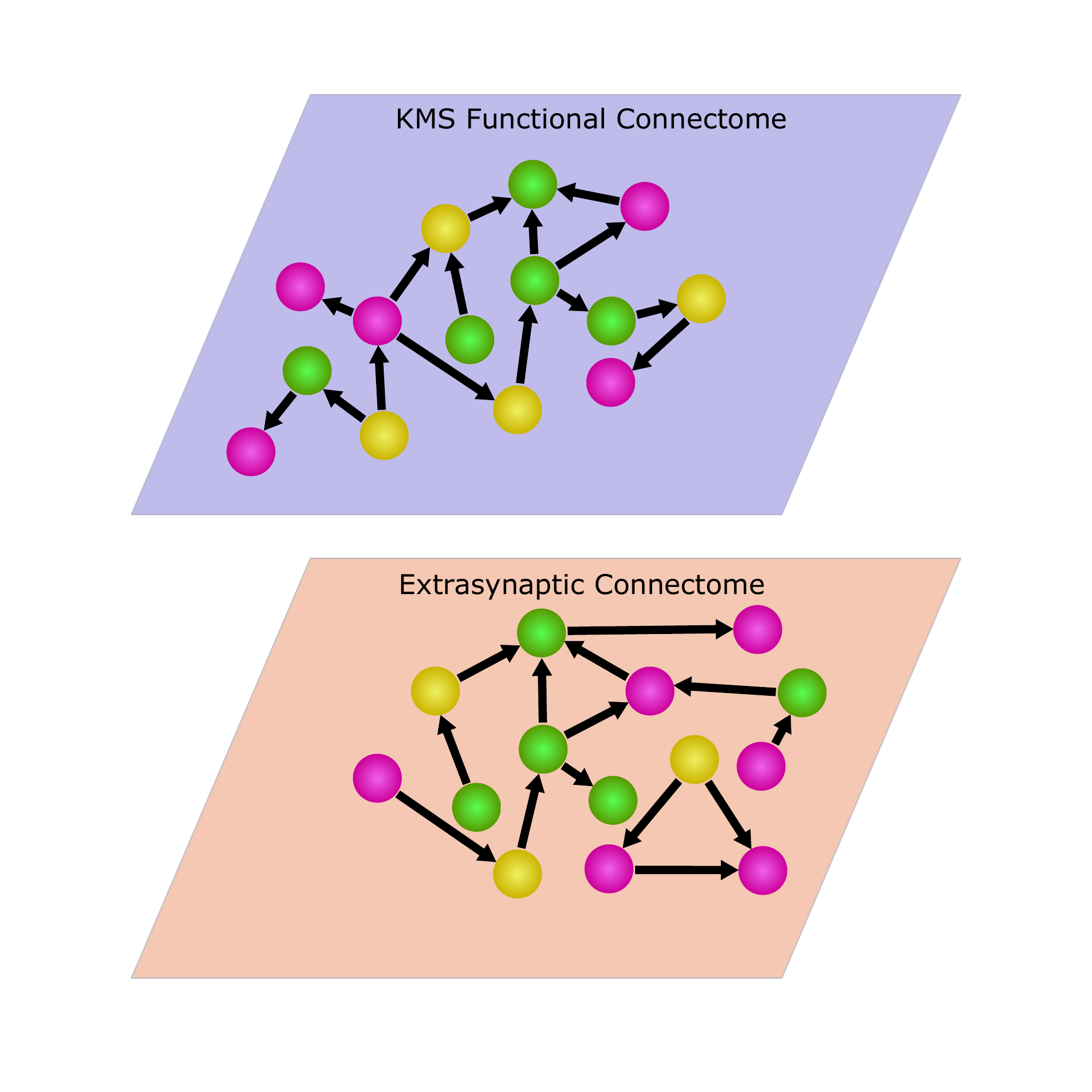}
  \caption{ \textbf{Functional multiplex framework integrating synaptic and extrasynaptic communication}. Schematic of the two-layer network combining the structure-informed functional connectome (SIFC), derived from synaptic wiring using the KMS framework, with the extrasynaptic neuropeptidergic connectome. Directed edges represent probabilistic information flow that may arise from direct or multi-step interactions. Nodes are colored by neuronal class (sensory: yellow, interneuron: green, motor: pink). This multiplex representation provides a common functional reference for comparing fast synaptic and diffusive extrasynaptic communication.}
  \label{fig:multiplex}
\end{figure}

\subsection{\label{sec:circuits}Functional Multiplex Network}

We next constructed a two-layer multiplex network combining SIFC with an independently
mapped extrasynaptic connectome (Fig.~\ref{fig:full_image}). The extrasynaptic layer,
derived from neuropeptide-receptor expression data~\cite{ripoll2023neuropeptidergic},
comprises 302 neurons and 54,035 directed edges representing potential diffusive signaling
pathways.

To identify functionally relevant synaptic interactions, we compared KMS-derived edge
weights against a degree-preserving null model (see SI). This approach isolates
contributions arising from network topology rather than node degree alone, and edges were
classified as significant when their weights exceeded the null distribution ($p < 0.05$).
The key insight is that extrasynaptic edges co-occurring with topology-constrained SIFC
connections are likely shaped by the same underlying wiring, whereas those co-occurring
with topology-insensitive connections --- or absent from the SIFC altogether --- reflect
signaling that is independent of synaptic architecture. We thus compared significant and
non-significant SIFC edges with extrasynaptic connections to define four communication
regimes, which characterize functional communication modes rather than intrinsic neuron
classes (see SI for classification procedures and statistical analyses):

\begin{enumerate}
    \item \textbf{Topology-dependent extrasynaptic regime} --- extrasynaptic edges
    overlapping with significant SIFC connections, implying dependence on synaptic
    topology.
    \item \textbf{Topology-resilient extrasynaptic regime} --- extrasynaptic edges
    overlapping with non-significant SIFC connections, implying independence from detailed
    synaptic structure.
    \item \textbf{Purely extrasynaptic regime} --- extrasynaptic edges absent from the
    SIFC, representing communication pathways entirely outside synaptic topology.
    \item \textbf{Purely synaptic regime} --- significant SIFC edges absent from the
    extrasynaptic connectome, representing fast, point-to-point communication channels.
\end{enumerate}

\subsubsection{\label{sec:dependent}Topology-dependent regime: Reinforcement of synaptic
motor circuits}

\begin{figure}[!h]
  \centering
  \includegraphics[width=\linewidth]{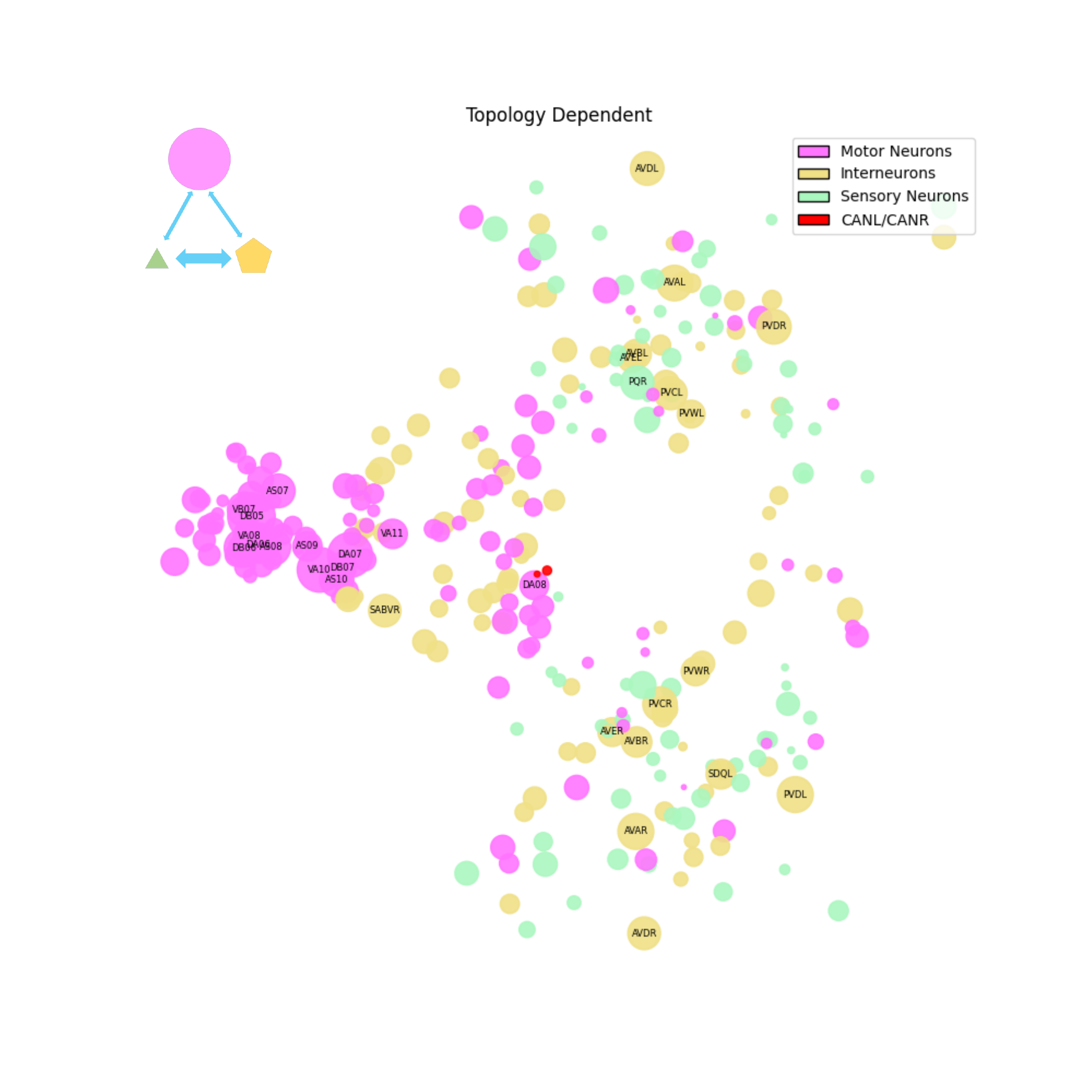}
  \caption{\textbf{Topology-dependent extrasynaptic regime reinforcing synaptic circuits}.
  Network representation of extrasynaptic connections that overlap with significant
  structure-informed functional connectivity. Node size reflects degree within this regime,
  with labels shown for the most connected neurons. This subnetwork is enriched for
  locomotion-related motor neurons and closely mirrors synaptic communication structure,
  showing that these extrasynaptic pathways are constrained by and reinforce underlying
  synaptic topology.}
  \label{fig:dep}
\end{figure}

The topology-dependent regime comprises 5,254 edges and closely mirrors the structure of
the SIFC, with nodes enriched for motor neurons associated with locomotion, including DB,
DA, VA, and AS classes (Fig.~\ref{fig:dep}). Highly connected nodes exhibit a strong
in-degree bias, revealing convergence of signaling onto key motor hubs. Neurons such as
DB05, DA07, VA10, and AS08 are among the most connected and are known to coordinate
locomotion, while sensory neurons such as PVD contribute by linking external stimuli to
motor circuits~\cite{wormatlas2023,chatzigeorgiou2010specific}. Conversely, neurons such
as VB07, AVBL, and AVBR exhibit outgoing signaling dominance and play a complementary role
in locomotor coordination~\cite{wormatlas2023,zhen2015}. This structural alignment with
the SIFC implies that these extrasynaptic pathways are constrained by synaptic topology
and, rather than forming independent circuits, reinforce existing synaptic communication
patterns. By providing parallel signaling routes, this regime may enhance motor pathway
reliability and increase robustness to perturbations in synaptic connectivity.

\subsubsection{\label{sec:independent}Topology-resilient regime: Global modulation
decoupled from synaptic wiring}

\begin{figure}[!h]
  \centering
  \includegraphics[width=\linewidth]{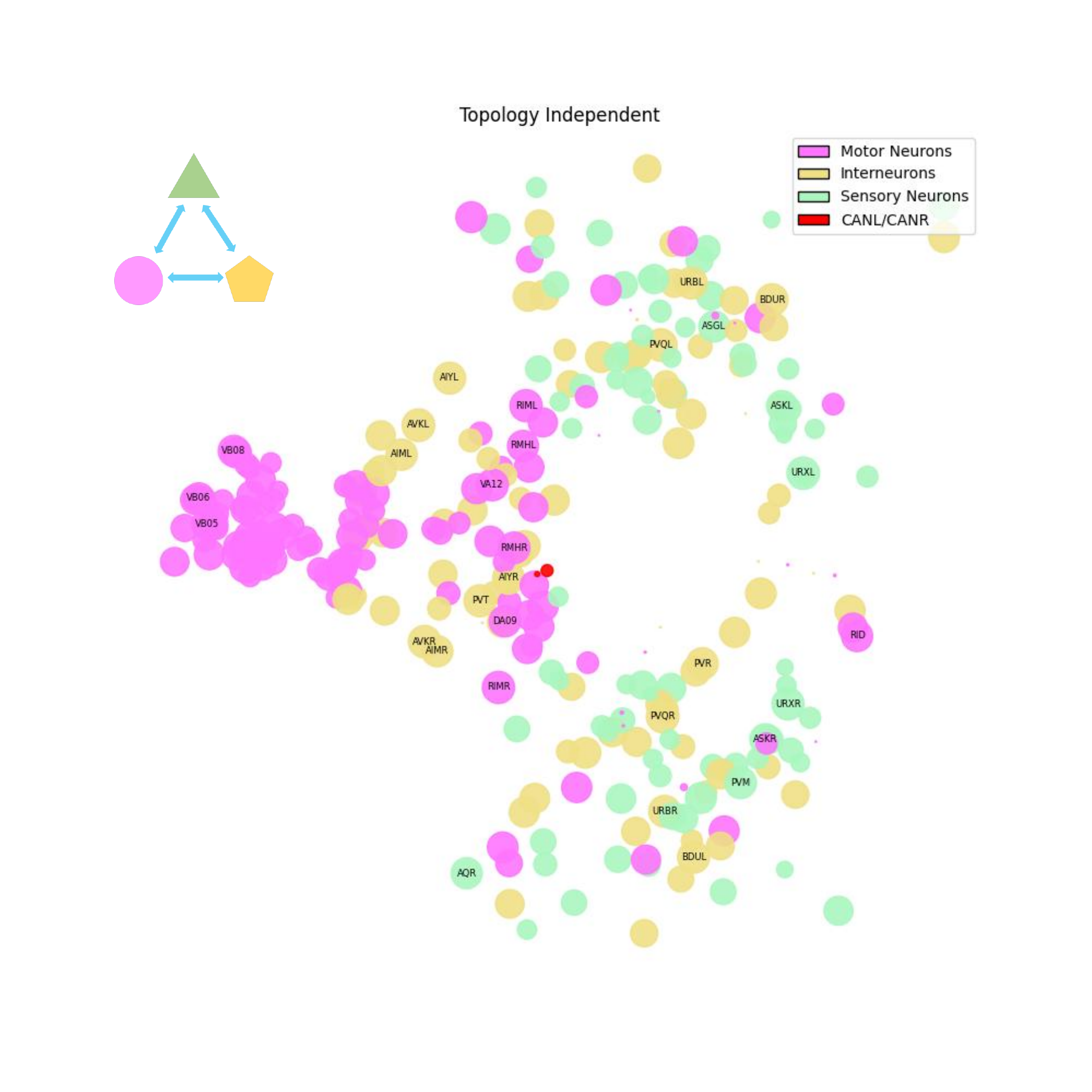}
  \caption{\textbf{Topology-resilient extrasynaptic regime supporting global modulation}.
  Extrasynaptic connections that persist independently of synaptic topology, remaining
  stable under degree-preserving randomization. Node size indicates connectivity within the
  regime. This network is dominated by interneurons involved in behavioral state regulation
  and global coordination, consistent with a distributed modulatory layer that operates
  independently of precise anatomical wiring.}
  \label{fig:indep}
\end{figure}

The topology-resilient regime comprises 42,022 edges and remains largely unchanged under
degree-preserving randomization of the synaptic connectome, implying weak dependence on
precise wiring architecture (Fig.~\ref{fig:indep}). In contrast to the
topology-dependent regime, nodes exhibit balanced in-degree and out-degree distributions.
Highly connected neurons include interneurons such as AVK and PVQ, known to regulate
behavioral state, arousal, and global
coordination~\cite{wormatlas2023,ji2023proprioceptive,marquina2024antagonism,
wadsworth1996neuroglia}, as well as RIM, DVA, and RIB, which are critical for bridging
the synaptic and extrasynaptic layers~\cite{bentley2016multilayer}. Sensory neurons such
as URX, which detect ambient oxygen levels and trigger avoidance
responses~\cite{wormatlas2023,cheung2005experience,gray2004oxygen}, are also prominent,
linking environmental signals to distributed network responses. The persistence of this
structure under randomization indicates a stable modulatory layer whose organization is
largely independent of the specific arrangement of synaptic edges, consistent with a role
in global state regulation --- including arousal, sleep, and behavioral transitions ---
that requires distributed coordination rather than rapid point-to-point transmission.

\subsubsection{\label{sec:pureextra}Purely extrasynaptic regime: Survival and homeostasis
beyond synaptic reach}

\begin{figure}[!h]
  \centering
  \includegraphics[width=\linewidth]{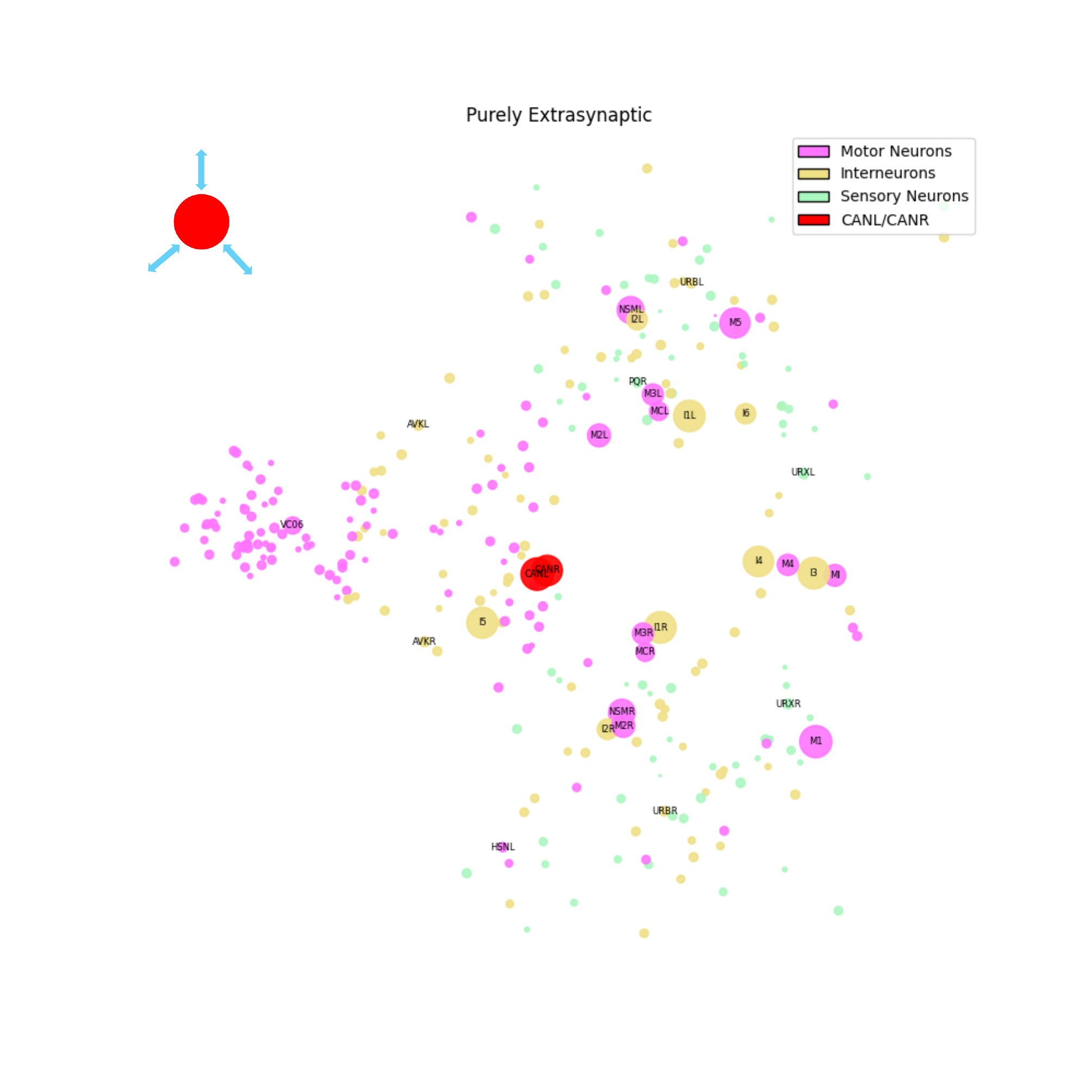}
  \caption{\textbf{Purely extrasynaptic regime supporting survival and homeostasis}.
  Extrasynaptic connections that do not overlap with synaptic functional connectivity,
  forming a distinct communication network outside synaptic topology. Node size reflects
  degree within this regime. Highly connected nodes include neurons essential for feeding,
  metabolism, and organismal viability, implying that critical life-support functions rely
  predominantly on extrasynaptic signaling.}
  \label{fig:pureextra}
\end{figure}

The purely extrasynaptic regime comprises 6,759 edges and operates entirely outside
synaptic topology: these connections are absent from both the SIFC and its randomized
counterparts, showing that they cannot be explained by synaptic structure alone
(Fig.~\ref{fig:pureextra}). This regime is enriched for neurons associated with survival
and homeostatic regulation. The most connected nodes include CANL/R, M1, I1, and M4 ---
neurons playing key roles in feeding, nutrient transport, and essential physiological
processes~\cite{wormatlas2023,trojanowski2014neural,bhatla2015light}. Strikingly, CANL,
CANR, and M4 are the three neurons individually required for the viability of
\emph{C.~elegans}~\cite{chien2019enigmatic}, yet all are sparsely connected in the
synaptic connectome. This contrast between limited synaptic wiring and high extrasynaptic
connectivity reveals that the most survival-critical neurons rely predominantly on
diffusive signaling, demonstrating that extrasynaptic communication can serve as a primary
--- rather than merely modulatory --- substrate for essential physiological function.

\subsubsection{\label{sec:puresyn}Purely synaptic regime: Fast sensorimotor processing
without extrasynaptic support}

\begin{figure}[!h]
  \centering
  \includegraphics[width=\linewidth]{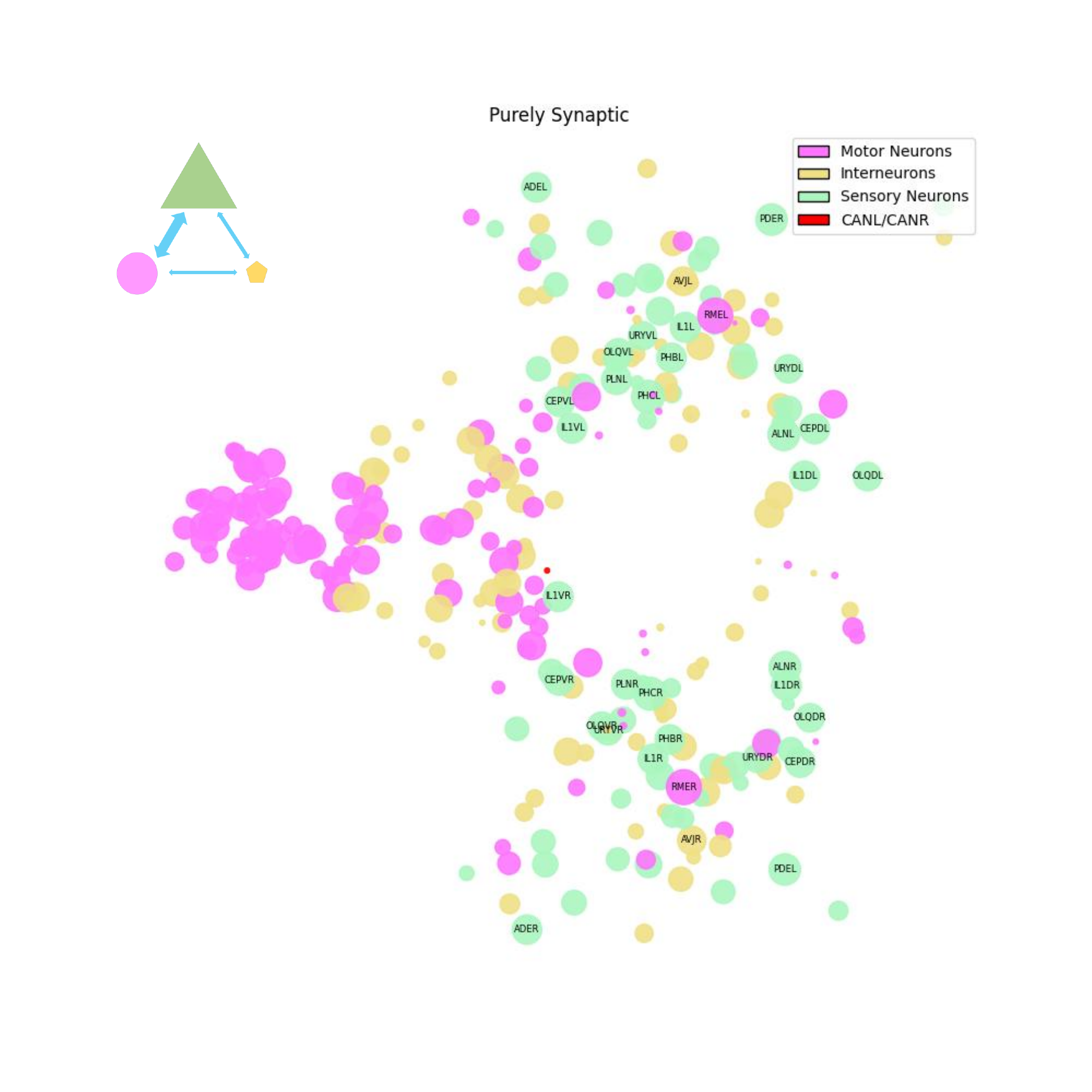}
  \caption{\textbf{Purely synaptic regime mediating rapid sensorimotor communication}.
  Functional synaptic connections not supported by extrasynaptic signaling. Node size
  represents connectivity within this subnetwork. This regime is enriched for sensory and
  motor neurons associated with fast behavioral responses, highlighting the specialization
  of synaptic wiring for rapid, low-latency information transfer.}
  \label{fig:puresyn}
\end{figure}

The purely synaptic regime comprises 30,950 edges, of which 2,649 are statistically
significant (Fig.~\ref{fig:puresyn}). This regime is enriched for sensory and motor
neurons involved in rapid behavioral responses. Highly connected nodes include RME motor
neurons and PHC sensory neurons, mediating head movement and environmental sensing
respectively, while AS10 and FLP --- associated with locomotion coordination and
mechanosensation --- concentrate the largest number of significant
connections~\cite{wormatlas2023,bhattacharya2019plasticity,altun2009high,
wang2024neurotransmitter,pereira2015cellular,tolstenkov2018functionally,kaplan1993}.
The dominance of sensorimotor neurons and the complete absence of extrasynaptic support
implies that this regime underlies fast, low-latency
communication~\cite{bentley2016multilayer,chatzigeorgiou2010specific}, consistent with
circuits where slower modulatory signaling would impair response precision.

The neuron-type composition across all four regimes is summarized in
Table~\ref{tab:connectivity_rank}, which reports the number of neurons with low, medium,
and high connectivity in each subgraph. Motor neurons dominate the topology-dependent
regime, interneurons are prominent in the topology-resilient regime, CAN neurons are
exclusively highly connected in the purely extrasynaptic regime, and sensory neurons are
most strongly represented in the purely synaptic regime --- a pattern that quantitatively
reinforces the functional specialization described above.

\begin{table}[H]
\caption{Number of neurons with low, medium, and high connectivity (by degree tercile
within each subgraph) across the four communication regimes, broken down by neuron type.
CAN neurons appear as highly connected only in the purely extrasynaptic regime, motor
neurons predominate in the topology-dependent regime, and sensory neurons are most
strongly represented in the purely synaptic regime.}
\medskip
\label{tab:connectivity_rank}
\begin{ruledtabular}
\begin{tabular}{lcccc}
\textbf{type} & \textbf{low} & \textbf{medium} & \textbf{high} & \textbf{graph} \\
\hline
INTERNEURONS    & 13 & 48 & 36 & dependent \\
MOTOR NEURONS   & 42 & 32 & 49 & dependent \\
SENSORY NEURONS & 45 & 25 & 10 & dependent \\
CANL/R          &  2 &  0 &  0 & dependent \\
INTERNEURONS    & 22 & 36 & 39 & independent \\
MOTOR NEURONS   & 44 & 31 & 48 & independent \\
SENSORY NEURONS & 33 & 33 & 14 & independent \\
CANL/R          &  2 &  0 &  0 & independent \\
INTERNEURONS    & 29 & 32 & 36 & purely extrasyn \\
MOTOR NEURONS   & 38 & 43 & 42 & purely extrasyn \\
SENSORY NEURONS & 47 & 22 & 11 & purely extrasyn \\
CANL/R          &  0 &  0 &  2 & purely extrasyn \\
INTERNEURONS    & 48 & 26 & 23 & purely synaptic \\
MOTOR NEURONS   & 42 & 40 & 41 & purely synaptic \\
SENSORY NEURONS & 10 & 33 & 37 & purely synaptic \\
CANL/R          &  2 &  0 &  0 & purely synaptic \\
\end{tabular}
\end{ruledtabular}
\end{table}

\section{\label{sec:discussion}Discussion and Conclusion}

This work establishes a unified functional framework linking synaptic and extrasynaptic
communication in the \emph{C.~elegans} nervous system. By converting the synaptic
connectome into a probabilistic map of information flow and using it as a reference frame
against which the neuropeptidergic connectome is compared, we identify four communication
regimes that partition neural function across complementary signaling modes --- a
principled multiplex organization that would be invisible to purely structural analyses.

A central result is that extrasynaptic signaling is not a diffuse extension of synaptic
wiring but a structured system with clearly distinct functional roles. Extrasynaptic
pathways segregate into a topology-dependent layer reinforcing synaptic motor circuits, a
topology-resilient modulatory layer supporting global coordination and behavioral state
regulation, and a purely extrasynaptic network sustaining survival and homeostasis.
Synaptic communication in turn forms a distinct regime specialized for rapid, low-latency
sensorimotor processing. Together, these regimes show that different signaling modalities
are optimized for specific computational demands --- speed, robustness, global regulation,
and physiological maintenance~\cite{bentley2016multilayer,atkinson2021ascaris} --- and
that slower mechanisms are not secondary to synaptic transmission but fulfill roles that
point-to-point wiring alone cannot.

The topology-dependent regime, enriched for locomotion-related motor neurons and closely
mirroring the SIFC, enhances reliability of motor pathways and may provide redundancy
under synaptic perturbation. The topology-resilient regime, dominated by behavioral-state
interneurons and stable under degree-preserving randomization, provides distributed
modulation decoupled from precise anatomical wiring. The purely extrasynaptic regime,
concentrated on survival-critical neurons such as CAN, M1, and M4 that are sparsely
connected synaptically, demonstrates that extrasynaptic signaling can serve as a primary
rather than merely modulatory substrate for essential physiology.

The present framework opens concrete directions for future development. Extending the KMS
formalism to multilayer networks would allow chemical and electrical synapses to be
assigned distinct propagation parameters, reflecting their different biophysical timescales
and yielding a finer-grained functional reference frame. In parallel, as the complete
monoamine connectome of \emph{C.~elegans} comes within
reach~\cite{bentley2016multilayer,flavell2013serotonin}, incorporating monoaminergic
alongside neuropeptidergic connectivity within the same multiplex framework would enable a
systematic comparison of multiple diffusive signaling modalities. More broadly, the KMS
equilibrium approach is deliberately complementary to dynamical models: rather than
simulating activity, it extracts the functional landscape encoded in the wiring, providing
the natural reference point for non-stationary extensions and a ready template for
cross-species analysis. By linking anatomical wiring to functional and modulatory dynamics
through a common thermodynamic representation, this work advances a unified view of how
diverse communication modes cooperate to sustain coherent brain function.

\bibliographystyle{naturemag}

\bibliography{refs} 

\noindent\rule{\linewidth}{0.4pt}
\section*{Supplementary information}
%% ---------------------------------------------------------------
%% Supplementary Information
%% Reset counters and prefix figures/equations/tables with "S"
%% ---------------------------------------------------------------
\setcounter{equation}{0}
\setcounter{figure}{0}
\setcounter{table}{0}
\setcounter{section}{0}
\renewcommand{\theequation}{S\arabic{equation}}
\renewcommand{\thefigure}{S\arabic{figure}}
\renewcommand{\thetable}{S\arabic{table}}

%% ---------------------------------------------------------------

\subsection*{S1.\; KMS functional connectivity and equilibrium interpretation}

The Kubo-Martin-Schwinger (KMS) formalism provides a rigorous characterization of
thermal equilibrium in systems with infinitely many degrees of
freedom~\cite{haag1967quantum}. A system is in thermal equilibrium at temperature $T$
when its correlation functions between observables satisfy an analytic condition relating
forward and time-reversed observables via a shift in complex time determined by the
inverse temperature $\beta = 1/(k_B T)$.

Recent work extends the KMS formalism to directed networks by modeling information flow
using graph algebras~\cite{raeburn2005graph, moutuou2025kubo, moutuou2025brain}. In this
setting $\beta$ acts as a propagation scale parameter controlling the relative
contribution of paths of different lengths: large $\beta$ emphasizes short, direct paths,
whereas smaller $\beta$ allows longer and recurrent pathways to contribute. This provides
a principled link between equilibrium statistical physics and probabilistic information
flow on networks.

\subsection*{S2.\; Graph algebra formulation and neural emittance}

The \emph{C.~elegans} connectome is modeled as a directed network $G=(V,E)$ comprising
302 neurons and 16,857 directed edges obtained by combining the chemical synapses and gap
junctions provided in the datasets by Varshney~\emph{et al.} and Cook~\emph{et
al.}~\cite{varshney2011structural, cook2019connectome}. Let $\mathbf{A}$ denote the
adjacency matrix.

The KMS construction assigns to each neuron $v$ a state $\mathbf{x}^{v|\beta}$,
representing the distribution of signals originating from $v$ under equilibrium
propagation. For target neuron $u$, the value of $\mathbf{x}^{v|\beta}$ is given by
\begin{equation}
\mathbf{x}^{v|\beta}_{u}
=
\frac{1}{Z^{\beta}_{v}}
\left(
\sum_{\mu:\,v\to u} e^{-\beta |\mu|}
\right),
\label{eq:SI_emittance}
\end{equation}
where the sum is over all directed walks $\mu$ of any length from $v$ to $u$, $|\mu|$ is
the number of synapses traversed, and $Z^{\beta}_{v}$ is a partition function quantifying
the total communication capacity of $v$ at inverse temperature $\beta$. Because the graph
contains cycles and self-loops, this series converges only for $\beta > \beta_c$, where
\begin{equation}
e^{-\beta_{c}} = \frac{1}{r(\mathbf{A})}, \qquad \beta_{c} = \log r(\mathbf{A}),
\end{equation}
with $r(\mathbf{A})$ the spectral radius of $\mathbf{A}$. In this regime the emittance
admits the resolvent form
\begin{equation}
\mathbf{x}^{v|\beta}_{u} = \frac{1}{Z^{\beta}_{v}}
\left( \mathbf{I} - e^{-\beta}\,\mathbf{A} \right)^{-1}_{uv},
\end{equation}
which captures both direct and higher-order pathways, integrating local and global
contributions to functional connectivity.

\subsection*{S3.\; Mixed KMS states and entropy}

Network-level equilibrium patterns are obtained by mixing neuron-specific KMS states
\begin{equation}
\mathbf{x} = \sum_{v} p_v \, \mathbf{x}^{v|\beta},
\end{equation}
where $\{p_v\}$ is a probability distribution over neurons. To quantify the dispersion of
information flow we compute the Shannon entropy of the resulting distribution. Higher
entropy indicates more distributed communication, whereas lower entropy reflects
concentration along a smaller set of pathways.

\subsection*{S4.\; Selection of functional inverse temperature}

\begin{figure*}[htbp]
\centering
\includegraphics[width=0.6\textwidth]{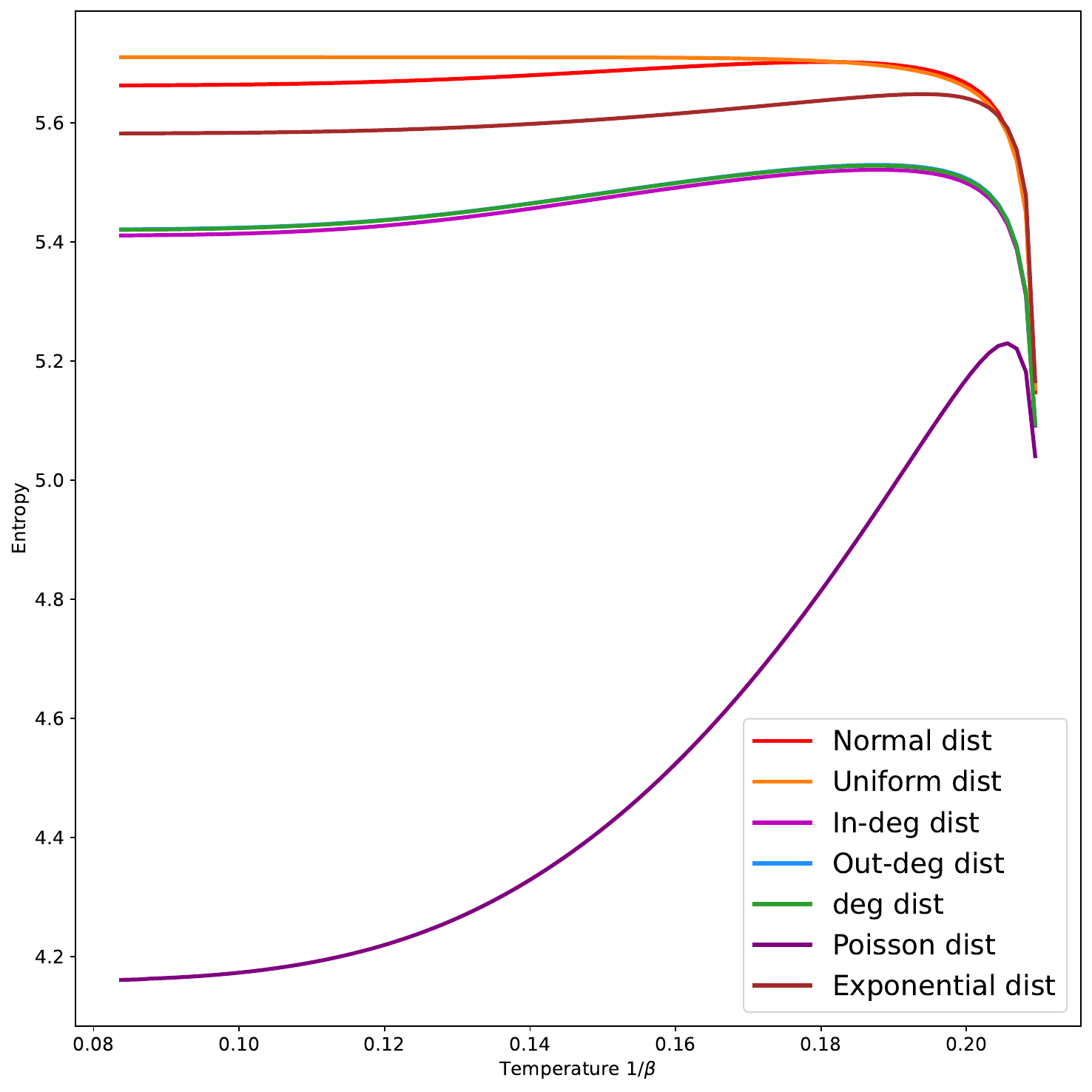}
\caption{\textbf{Entropy of the mixed emittance distribution as a function of inverse
temperature.} Shannon entropy of the mixed KMS state as a function of $1/\beta$ for
seven neuron-weighting schemes (in-degree, out-degree, total degree, normal, uniform,
Poisson, and exponential distributions). The selected value $\beta = 5.103$ maximizes
entropy across all weighting conditions while remaining within the convergence regime,
corresponding to maximally distributed yet stable information flow.}
\label{fig:SI_entropy}
\end{figure*}

Entropy was evaluated across a range of $\beta$ using multiple neuron-weighting schemes
(in-degree, out-degree, total degree, and standard reference distributions). We selected
$\beta = 5.103$, which maximizes entropy across all weighting schemes while remaining
within the convergence regime (Fig.~\ref{fig:SI_entropy}), corresponding to a maximally
distributed yet stable communication regime.

\subsection*{S5.\; Dataset description}

The synaptic connectome was assembled by merging two datasets: the physical connectome of
\emph{C.~elegans} reported by Varshney~\emph{et al.}~\cite{varshney2011structural} and
the whole-animal connectomes of both sexes published by Cook~\emph{et
al.}~\cite{cook2019connectome}. All edges from both datasets were combined to produce a
comprehensive connectome containing 302 neurons and 16,857 directed edges, which served
as input for the KMS functional connectivity computation. The neuropeptidergic connectome
published by Ripoll-S\'{a}nchez~\emph{et al.}~\cite{ripoll2023neuropeptidergic} was used
as the extrasynaptic layer for comparison.

\subsection*{S6.\; Null model and statistical significance}

To assess the dependence of functional interactions on synaptic topology, we generated
5,000 degree-preserving randomized networks. Each null network preserves the in-degree
and out-degree of every neuron while randomizing edge
placement~\cite{fosdick2016configuring}. For each neuron pair, the observed KMS emittance
value was compared to the null distribution; edges exceeding the 95th percentile ($p <
0.05$) were classified as significant, isolating interactions specifically constrained by
network topology rather than degree alone.

\subsection*{S7.\; Subgraph classification and network analysis}

Functional subgraphs were defined by intersecting KMS-derived edges with the extrasynaptic
connectome:
\begin{itemize}
  \item \textbf{Topology-dependent extrasynaptic} --- extrasynaptic edges overlapping
        with significant KMS edges.
  \item \textbf{Topology-resilient extrasynaptic} --- extrasynaptic edges overlapping
        with non-significant KMS edges.
  \item \textbf{Purely extrasynaptic} --- extrasynaptic edges absent from KMS
        connectivity entirely.
  \item \textbf{Purely synaptic} --- significant KMS edges without extrasynaptic
        support.
\end{itemize}

These subgraphs define functional communication regimes rather than intrinsic neuron
classes. For each subgraph, network statistics including degree, weighted degree, and
neuron-type composition were computed. Motor neurons dominated the topology-dependent
network, interneurons were prominent in the topology-resilient network, survival-critical
neurons (CAN, M4, I1) were concentrated in the purely extrasynaptic subgraph, and sensory
neurons were most strongly represented in the purely synaptic subgraph, consistent with
its role in rapid reflexive behavior.

\subsection*{S8.\; Topology-dependent regime: reinforcement of synaptic motor circuits}

\begin{figure*}[htbp]
\centering
\includegraphics[width=0.6\textwidth]{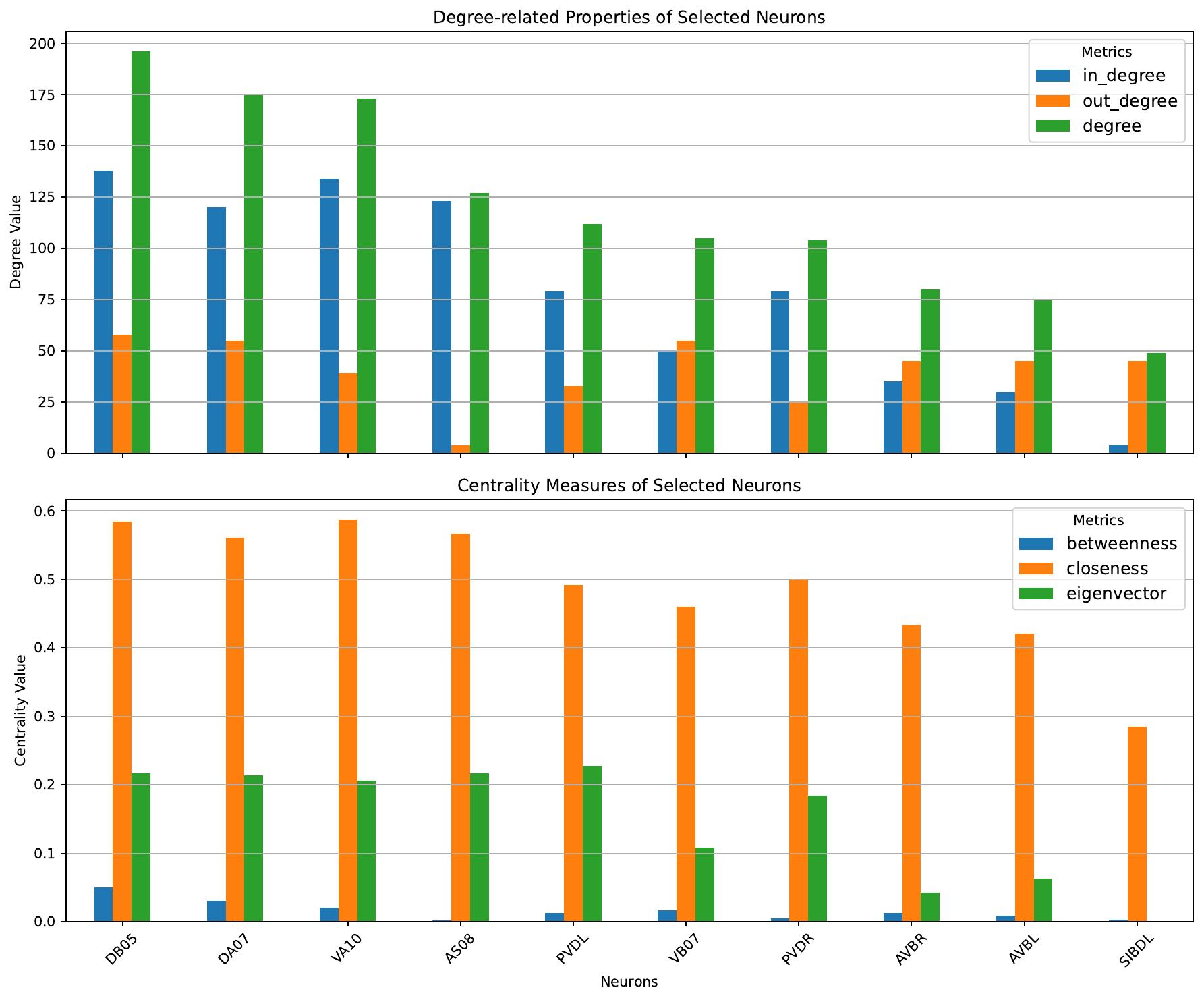}
\caption{\textbf{Connectivity and centrality measures for the topology-dependent
extrasynaptic subgraph.} Degree (in, out, total) and centrality measures (betweenness,
closeness, eigenvector) for the highest-degree neurons within this regime. In-degree
dominance is visible for key locomotion motor neurons (DB05, DA07, VA10, AS08), consistent
with convergence of signaling onto motor hubs, while VB07, AVBL, and AVBR show higher
out-degree reflecting broadcast roles.}
\label{fig:SI_cde}
\end{figure*}

This regime is enriched for locomotion-related neurons, including DB05, DA07, VA10, and
AS08 (motor; locomotion) and PVDL/R (sensory;
mechanosensation)~\cite{wormatlas2023,chatzigeorgiou2010specific}. High-degree nodes show
in-degree dominance, consistent with convergence of signaling onto motor hubs. Other
nodes, such as VB07, AVBL, and AVBR (motor; locomotion), exhibit higher out-degree,
indicating broadcast roles within locomotor circuits~\cite{wormatlas2023,zhen2015}. This
directional organization, together with the close structural correspondence to the SIFC,
indicates that these extrasynaptic pathways reinforce synaptic motor communication and may
enhance circuit robustness under perturbation (Fig.~\ref{fig:SI_cde}).

\subsection*{S9.\; Topology-resilient regime: global modulation decoupled from synaptic
wiring}

\begin{figure*}[htbp]
\centering
\includegraphics[width=0.6\textwidth]{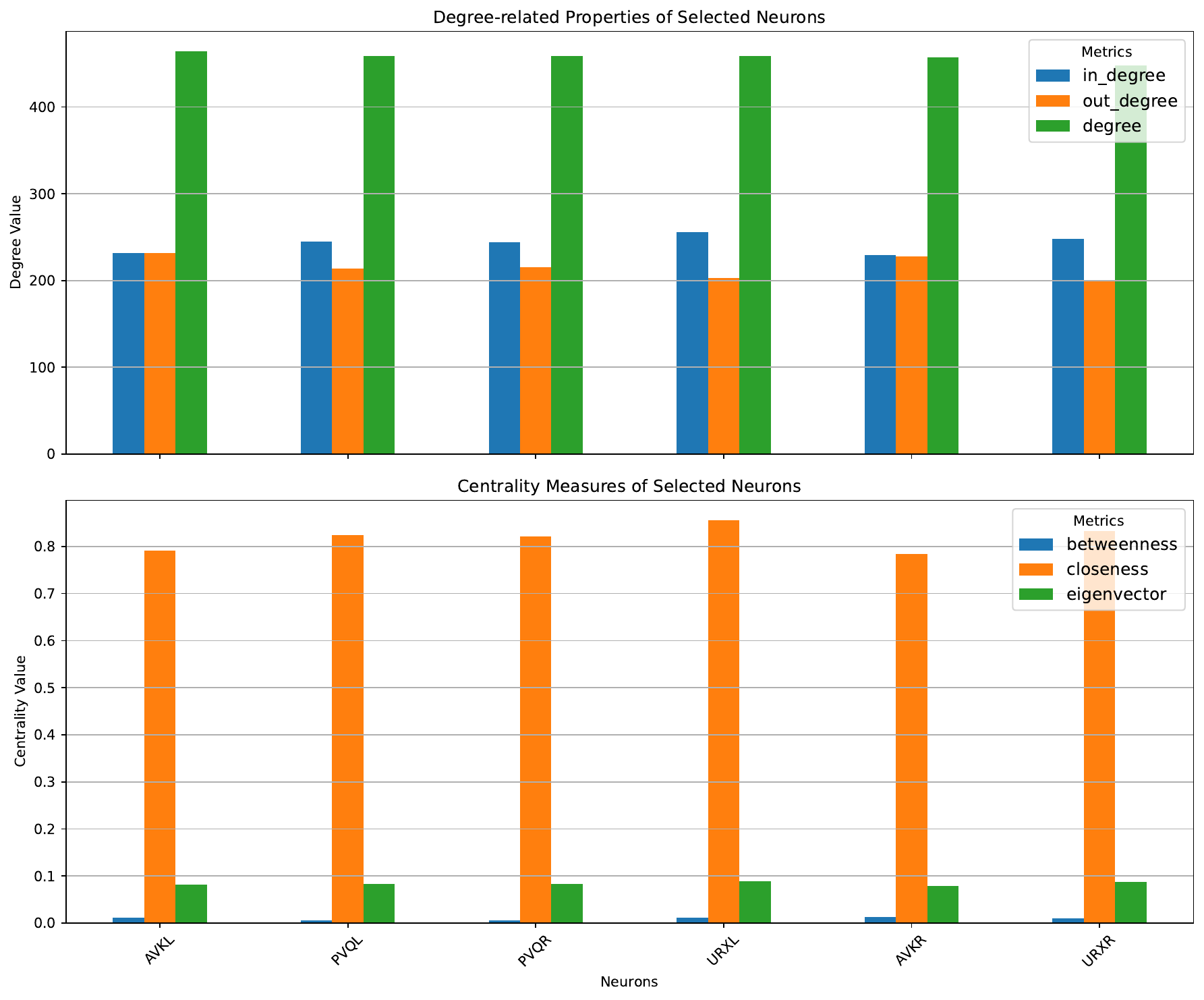}
\caption{\textbf{Connectivity and centrality measures for the topology-resilient
extrasynaptic subgraph.} Degree and centrality measures for representative highly
connected neurons in this regime. Balanced in- and out-degree is characteristic across
nodes, reflecting the distributed modulatory role of this layer. AVKL/AVKR, PVQL/PVQR,
and URXL/URXR form a symmetric, highly connected core.}
\label{fig:SI_cie}
\end{figure*}

This regime is dominated by interneurons such as AVK (AVKL/AVKR; 232 in- and out-edges
each for AVKL) and PVQ, known for coordination roles, together with RIM, RIB, and DVA,
which occupy central cross-layer positions bridging the synaptic and extrasynaptic
layers~\cite{wormatlas2023,ji2023proprioceptive,marquina2024antagonism,wadsworth1996neuroglia,
bentley2016multilayer}. URX (sensory; oxygen sensing) is also prominent, linking
environmental signals to distributed network
responses~\cite{wormatlas2023,cheung2005experience,gray2004oxygen,zimmer2009neurons}.
Connectivity is typically balanced between in- and out-degree. In contrast, pharyngeal
neurons (e.g., M3, I2R) are absent, confirming that this regime is organized around a
broad synaptic--extrasynaptic interface rather than local
microcircuits~\cite{wormatlas2023,dent1997avr}. The persistence of this structure under
degree-preserving randomization indicates a stable modulatory layer with weak dependence
on precise wiring (Fig.~\ref{fig:SI_cie}).

\subsection*{S10.\; Purely extrasynaptic regime: survival and homeostasis beyond synaptic
reach}

\begin{figure*}[htbp]
\centering
\includegraphics[width=0.6\textwidth]{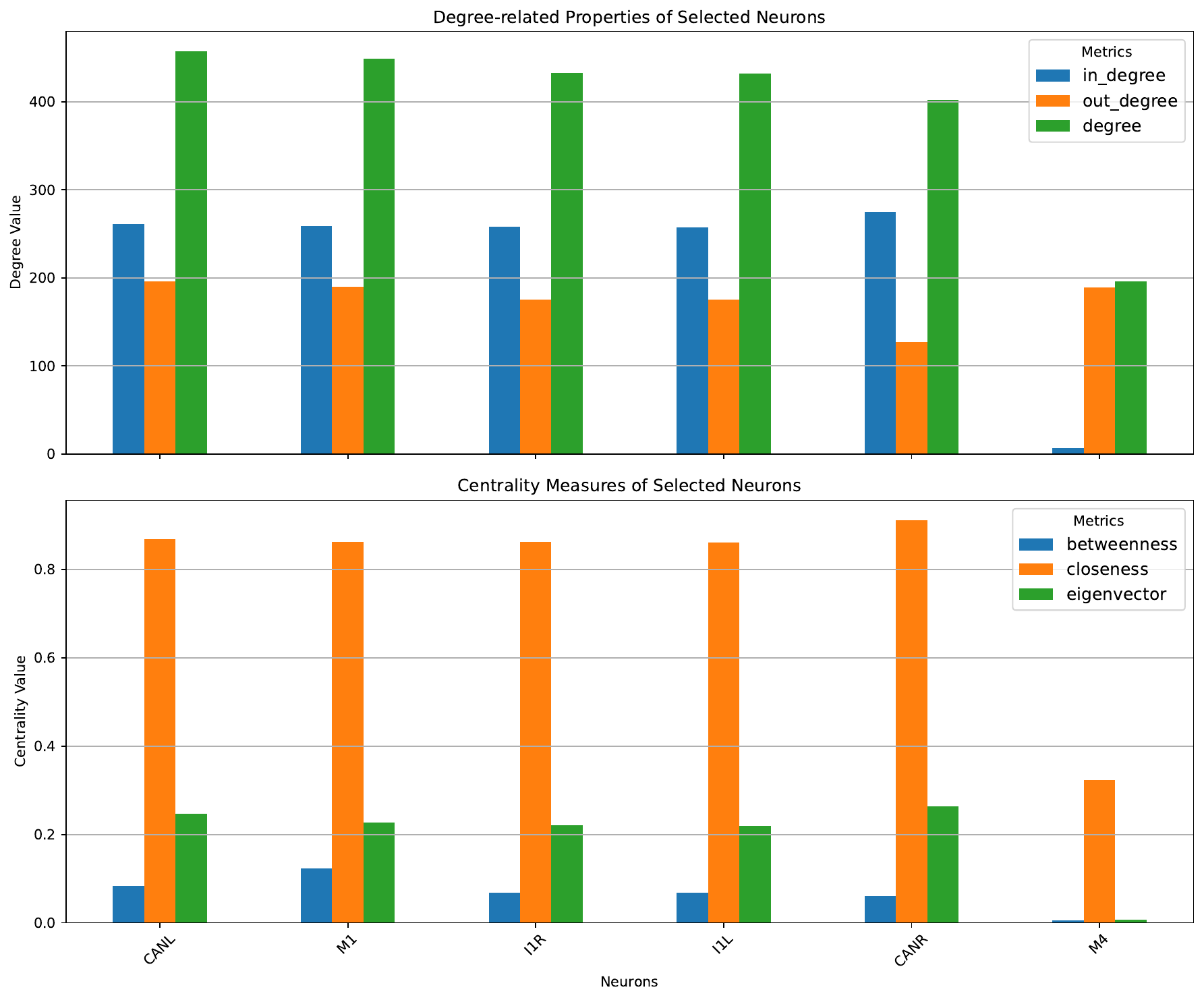}
\caption{\textbf{Connectivity and centrality measures for the purely extrasynaptic
subgraph.} Degree and centrality measures for the most connected neurons within this
regime, including CANL, M1, I1R, I1L, CANR, and M4. These neurons are sparsely connected
in the synaptic connectome but highly connected extrasynaptically, highlighting their
reliance on diffusive signaling for survival-critical and homeostatic functions.}
\label{fig:SI_peg}
\end{figure*}

This regime is enriched for neurons linked to survival and homeostasis, including CANL
(457 connections), M1 (449), I1L/R, CANR (402), and M4 (196). I1 and M1 regulate feeding
by coupling sensory input to pharyngeal
output~\cite{wormatlas2023,trojanowski2014neural,bhatla2015light}, while M4 is required
for peristalsis and effective food transport~\cite{wormatlas2023,trojanowski2014neural,
song2012serotonin}. These neurons are sparsely connected in the synaptic connectome but
highly connected extrasynaptically; key members (CANL/R, M4) are individually required
for organismal viability~\cite{chien2019enigmatic}. This contrast indicates that critical
physiological functions in this regime rely predominantly on extrasynaptic communication,
which acts as a primary --- not merely modulatory --- signaling substrate
(Fig.~\ref{fig:SI_peg}).

\subsection*{S11.\; Purely synaptic regime: fast sensorimotor processing without
extrasynaptic support}

\begin{figure*}[htbp]
\centering
\includegraphics[width=0.6\textwidth]{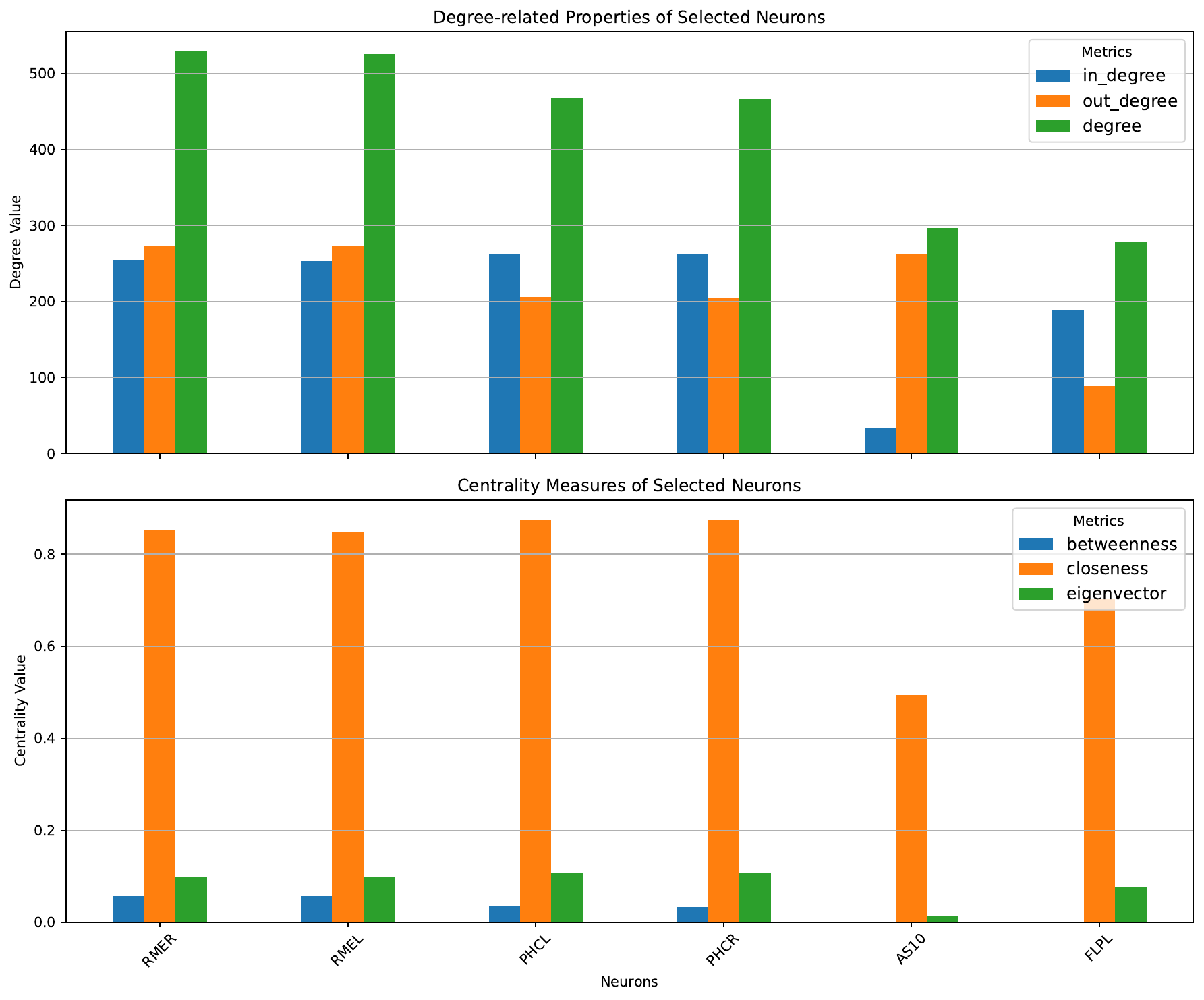}
\caption{\textbf{Connectivity and centrality measures for the purely synaptic subgraph.}
Degree and centrality measures for the highest-degree neurons within this regime,
including RMER, RMEL, PHCL, PHCR, AS10, and FLPL. Although RMEL/RMER and PHCL/PHCR have
high total degree, most of their connections are non-significant; AS10 and FLPL
concentrate the largest number of topology-specific significant connections.}
\label{fig:SI_psg}
\end{figure*}

This regime captures significant KMS-derived interactions without extrasynaptic support
(2,649 significant vs.\ 28,301 non-significant connections). It is enriched for
sensorimotor neurons: RME (motor; head movement), PHC (sensory; environmental sensing),
AS (motor; locomotion coordination), and FLP (sensory;
mechanosensation)~\cite{kaplan1993,wormatlas2023,bhattacharya2019plasticity,altun2009high,
wang2024neurotransmitter,pereira2015cellular,tolstenkov2018functionally}. RMEL/RMER and
PHCL/PHCR are major hubs by total degree, although many of their connections are
non-significant, indicating that high degree does not necessarily reflect
topology-specific enrichment. By contrast, AS10 and FLPL concentrate the largest number
of significant connections. Overall, this regime supports fast, low-latency sensorimotor
communication where direct synaptic transmission is essential (Fig.~\ref{fig:SI_psg}).

%% ---------------------------------------------------------------
\subsection*{Supplementary Figures}
%% ---------------------------------------------------------------

\end{document}